\documentclass[
 reprint,nofootinbib,
 amsmath,amssymb,
 aps,
]{revtex4-2}

\usepackage{graphicx}
\usepackage{dcolumn}
\usepackage{bm}
\usepackage{xcolor}
\usepackage[colorinlistoftodos]{todonotes}

\usepackage{soul}

\begin{document}

\preprint{APS/123-QED}

\title{Learning Heat Transport Kernels Using a Nonlocal Heat Transport Theory-Informed Neural Network}

\author{Mufei Luo${}^{1}$, Charles Heaton${}^{1}$, Yizhen Wang${}^{1}$, Daniel Plummer${}^{1}$, Mila Fitzgerald${}^{1}$, Francesco Miniati${}^{2}$, Sam M. Vinko${}^{1}$ and Gianluca Gregori${}^{1}$}

\affiliation{%
 ${}^1$ Department of Physics, University of Oxford, Parks Road, Oxford OX1 3PU, UK
}%
\affiliation{%
 ${}^2$ Mach42, Robert Robinson Avenue, Oxford Science Park, Oxford, OX4 4GP, UK
}%

\begin{abstract}
We present a data-driven framework for the modeling of nonlocal heat transport in plasmas using a nonlocal theory informed-neural network trained on kinetic Particle-in-Cell simulations that span both local and nonlocal regimes. The model learns spatio-temporal heat flux kernels directly from simulation data, capturing dynamic transport behaviors beyond the reach of classical formulations. Unlike time-independent kernel models such as Luciani–Mora–Virmont and Schurtz–Nicolaï–Busquet models, our approach yields physically grounded, time-evolving kernels that adapt to varying plasma conditions. The resulting predictions show strong agreement with kinetic benchmarks across regimes. This offers a promising direction for data-driven modeling of nonlocal heat transport and contributes to a deeper understanding of plasma dynamics.
\end{abstract}

\maketitle

Understanding heat transport~\cite{Gianluca1} in plasmas is essential for a broad range of applications, from inertial confinement fusion~\cite{ICFoverreviw} to astrophysical systems~\cite{Yerger,Komarov}. 
Heat transport in hot plasma is characterised by the ratio of the temperature gradient length $L_T=T_e/\nabla T_e$, where $T_e$ is the electron temperature, to the electron mean free path $\lambda_{\rm free}=4 \pi\varepsilon_0^2 m_e^2 v_{\rm {th}}^4/(Z n_ee^4 \ln\Lambda)$~\cite{huba2023plasma}, where $\varepsilon_0$ is the vacuum permittivity, $m_e$ is the electron mass, $v_{th}=\sqrt{T_e/m_e}$ is the electron thermal velocity, $Z$ is the ion charge state, $n_e$ is the electron number density, $e$ is the elementary charge, and $\ln\Lambda$ is the Coulomb logarithm. In the local regime where $L_T/\lambda_{\rm free} \gg 500$, classical models such as the Spitzer–Härm (SH) formulation~\cite{SH} yield reliable predictions. However, as this ratio decreases, nonlocal effects emerge, invalidating local hydrodynamic closures~\cite{nonlocalclosure} and necessitating more advanced descriptions. Existing transport models include the Luciani–Mora–Virmont (LMV)~\cite{LMV}, Schurtz–Nicolaï–Busquet (SNB)~\cite{SNB}, and flux-limited schemes, attempt to extend classical theory by incorporating empirical or semi-analytic kernels. Yet these models generally assume temporal stationarity and rely on simplified or heuristic approximations, limiting their validity in regimes where the electron collision time approaches hydrodynamic time scales.

At the core of the popular nonlocal transport theories lies the heat flux kernel, which mediates the spatial coupling between local SH flux and its nonlocal contributions. In current analytic models, this kernel is often time-independent, thereby neglecting dynamic features that are critical for the accurate modeling in highly nonlocal regimes. This motivates the need for a more flexible framework—one capable of learning transport kernels directly from kinetic data, while retaining interpretability and physical structure.

Fully kinetic simulations, such as Vlasov-Fokker–Planck (VFP) and Particle-in-Cell (PIC) methods, offer high-fidelity descriptions of heat transport but vary greatly in computational cost. VFP solvers~\cite{bell,zhao}, while detailed, are often prohibitively expensive for large parameter studies. PIC simulations, when equipped with physically validated binary collision models~\cite{piccollision1, piccollision2}, offer a computationally efficient yet accurate means to capture nonlocal heat transport~\cite{collisonexample1,collisonexample2,collisonexample3}, bridging the gap between fidelity and scalability. In this work, we applied a 1D PIC code (OSIRIS~\cite{osiris}) to generate a large, structured dataset across local and nonlocal regimes, enabling data-driven construction of dynamic heat flux kernels via neural networks.

To investigate heat transport across a wide range of plasma conditions, we impose an initial electron temperature profile of the form
\begin{equation}
T_e(x) = \frac{2T_{e0}(R - 1)/(R + 1)}{1 + \exp(x/L)} + \frac{2T_{e0}}{R + 1},
\label{teprofile}
\end{equation}
where $T_{e0} = 1$ keV sets the central temperature, $R = T_{h}/T_{c}$ is the hot-to-cold temperature ratio, and \textcolor{black}{the steepness length $L$ determines the gradient scale length via $L_T = 2L(R + 1)/(R - 1)$}. The temperature profile described by Eq.~(\ref{teprofile}) features a hot region on the left and a cold region on the right, with the gradient localized at the center over a width of approximately $L$. The corresponding density profile is chosen as $n_e(x) \propto 1/T_e(x)$, consistent with isobaric heat conduction commonly observed in the conduction zone of laser-driven fusion plasmas~\cite{suxinghu1}. To control the degree of nonlocality, we vary the background electron density, thereby tuning the mean free path $\lambda_{\rm free}$ relative to $L_T$. This allows systematic exploration of transport behavior from the classical to the strongly nonlocal regime.

The initial condition is parameterized by the tuple $(R,\ L,\ n_{e0})$, where $R$ is the temperature ratio, $L$ determining the gradient scale length, and $n_{e0}$ the central electron density. Note that $R$ and $L$ are not entirely independent. We will discuss this further below. To ensure physically meaningful regimes and non-degenerate conditions, we consider electron densities ranging from $10^{19}$ to $10^{25}~\mathrm{cm^{-3}}$, corresponding to Fermi temperatures up to approximately $0.17$ keV. We assume a fully ionized, high-$Z$ plasma ($Z = 16$) to enhance collisionality and reduce relaxation times. The Coulomb logarithm $\rm ln\Lambda$, calculated at the box center, is held constant for each simulation. All simulations are performed in 1D geometry with spatial and temporal units normalized to the plasma wavenumber $k_{\rm pe}$ and frequency $\omega_{\rm pe}$. The box length is fixed at $L_{\rm box} \approx 1500$, with spatial resolution $\Delta x = 0.04$ and time step $\Delta t = 0.035$, satisfying Courant–Friedrichs–Lewy condition ($\Delta t / \Delta x \approx 0.9$) and resolving the Debye length ($\lambda_{\rm De,0} = 1.2 \Delta x$). Each cell contains 25,000 electron and 1,600 ion macroparticles. 
A total of 829 data points are sampled from the parameter space, covering a range of $L_T / \lambda_{\rm free}$ from 10 to 5000, spanning regimes from strongly nonlocal to classical heat transport.

\begin{figure}
    \centering
    \includegraphics[width=1\linewidth]{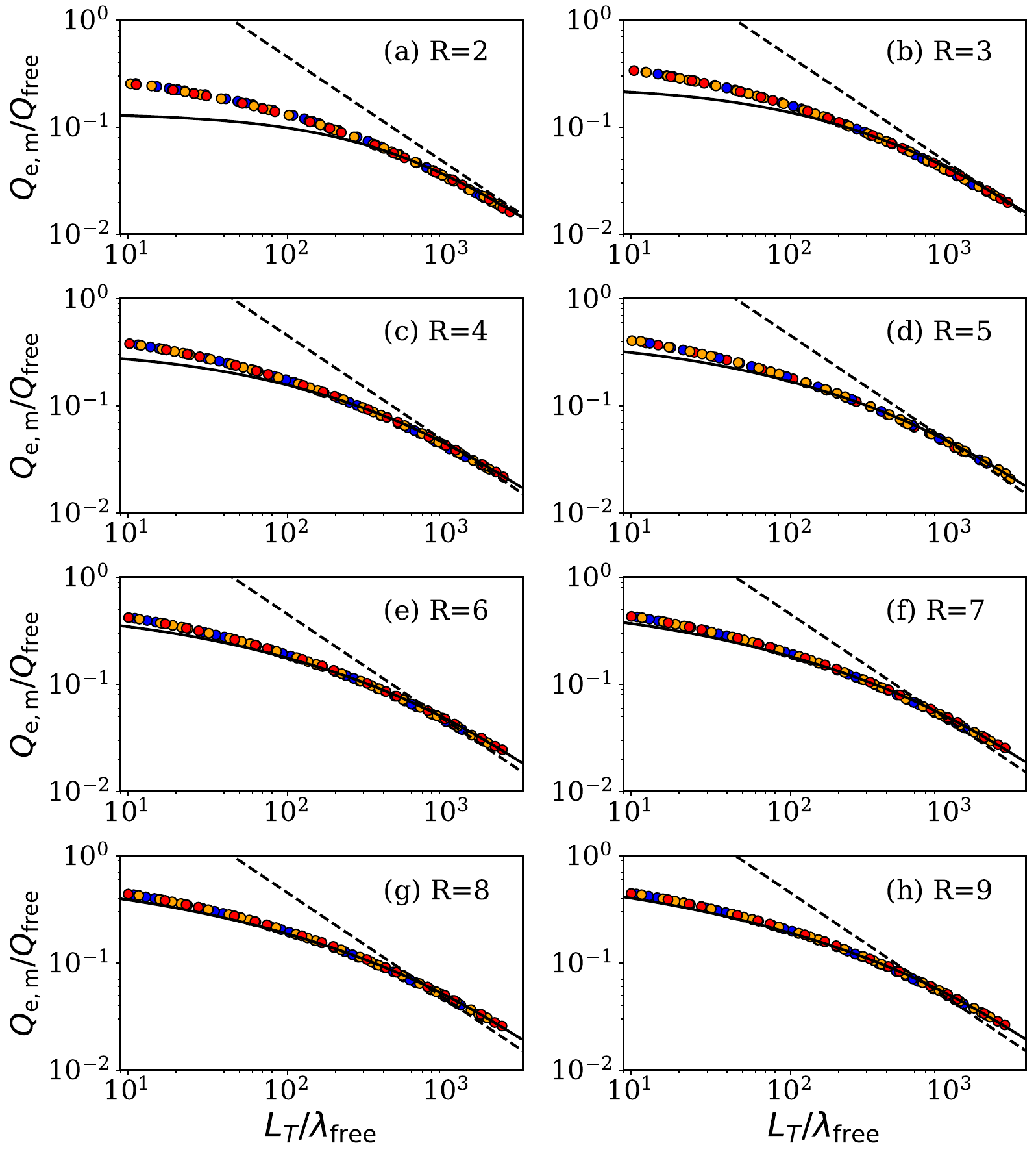}
    \caption{
Maximum normalized electron heat flux $Q_{\rm e,m}/Q_{\rm free}$ as a function of the initial local transport ratio $L_T/\lambda_{\rm free}$ evaluated at the center of the domain ($x = 0$). Each subplot corresponds to a different temperature ratio $R = T_{h}/T_{c}$, increasing from (a) $R=2$ to (h) $R=9$.  
In panel (a), the blue, orange, and red points correspond to steepness lengths of $L \approx 16$, $24$, and $32$, respectively. In panels (b)–(h), the steepness length $L$ associated with each color (as defined in panel (a)) is adjusted for each temperature ratio $R$ using the relation $L_T = 2L(R + 1)/(R - 1)$, so that the ratio $L_T/\lambda_{\rm free}$ at the domain center remains the same for one given central density $n_{e0}$ across all panels. This design enables systematic exploration of nonlocal transport behavior across a range of physical regimes. The black dashed and solid lines represent the predictions of the SH and LMV models, respectively.
}
    \label{picdata1}
\end{figure}

\begin{figure*}
    \centering
    \includegraphics[width=1\linewidth]{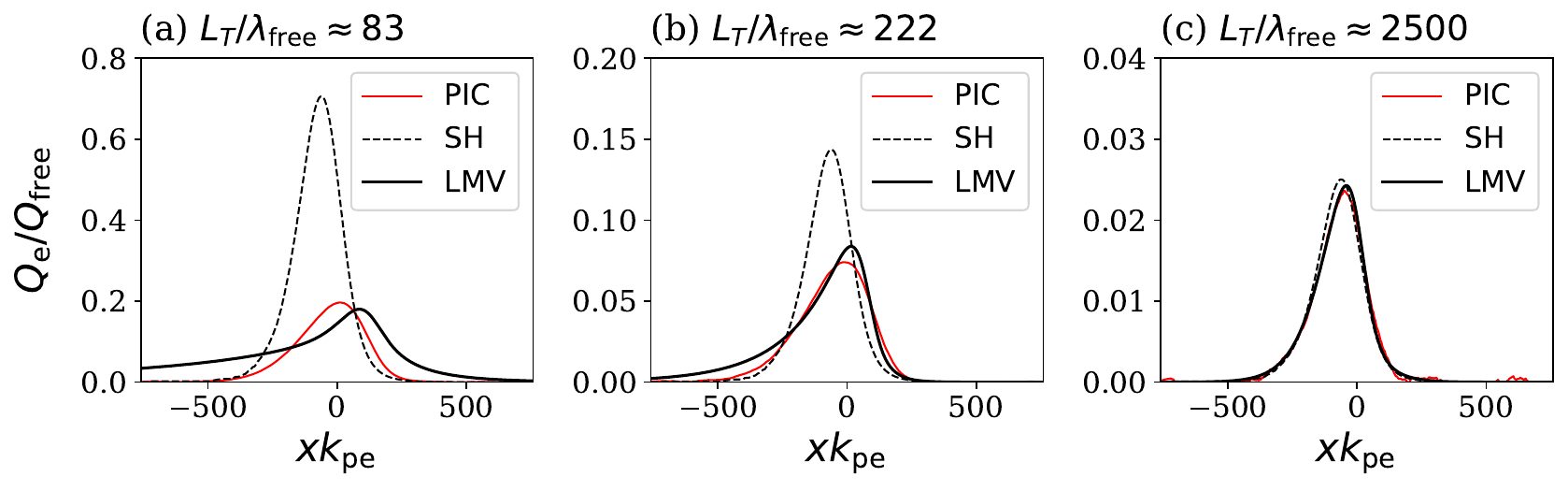}
    \caption{
Comparison of the spatial profiles of the saturated heat flux obtained from PIC simulations (red solid lines), SH theory (black dashed lines), and the LMV model (black solid lines) for different plasma densities. The temperature ratio is fixed at $R = 5$, and the gradient scale length is set to $L_T\sim 195$. From (a) to (c), the background electron density increases from $10^{21}\mathrm{cm}^{-3}$ to $10^{25}\mathrm{cm}^{-3}$, corresponding to $L_T/\lambda_{\rm free} \approx 83$, 222, and 2500, respectively.}
    \label{picdata2}
\end{figure*}

\textcolor{black}{The heat flux is calculated by $Q_e=m_e\int v^3f(v)dv$ as no current is driven, where $f(v)$ is the electron velocity distribution.} After noise filtering, the maximum electron heat flux $Q_{\rm e,m}$, normalized by the free-streaming flux ($Q_{\rm free} = n_{e0} T_{e0} v_{\rm th0}$, where $v_{\rm th0} = \sqrt{T_{e0}/m_e}$ is the electron thermal velocity at the center of the simulation box), is shown in Fig.~\ref{picdata1}. The $x$-axis represents the initial value of $L_T/\lambda_{\rm free}$, evaluated at the center of the domain ($x = 0$). In Figs.~\ref{picdata1}(a) through (h), the temperature ratio $R = T_{h}/T_{c}$—defined as the ratio of the hot-side to cold-side boundary temperatures—is varied from 2 to 9. 
In Fig.~\ref{picdata1}(a) with $R = 2$, the blue, orange, and red points correspond to steepness lengths of $L \approx 16$, $L \approx 24$, and $L \approx 32$, respectively. We avoid choosing excessively large values of $L$ to prevent heat flux from being driven near the boundaries.
From Fig.~\ref{picdata1}(b) to (h), as the temperature ratio $R$ increases, the steepness length $L$ associated with each color (as defined in Fig.~\ref{picdata1}(a)) is adjusted according to $L_T = 2L(R + 1)/(R - 1)$. For the given central density $n_{e0}$—and hence a given mean free path—this ensures that the ratio $L_T/\lambda_{\rm free}$ remains constant at the center of the domain.
It should be noted that all simulations are terminated at $t\omega_{\rm pe}=$2500 for their respective background densities. This duration is generally sufficient to saturate the transport process in all samples.

The black dashed lines in each subplot represent the classical local heat transport prediction of SH flux, given by $Q_{\rm SH}/Q_{\rm free} = a\, \lambda_{\rm SH} / L_T$, where $a = 128(Z + 0.24) / [3\pi(Z + 4.2)]$ and $\lambda_{\rm SH} = 3 \lambda_{\rm free} / \sqrt{\pi/2}$ is the classical Spitzer–Härm mean free path. To validate the accuracy of our data in the nonlocal regime, we also include the 1D nonlocal theoretical prediction as black solid lines. This LMV model follows the original formulation by Luciani and Mora~\cite{LMV}, in which the heat flux is expressed as

\begin{equation}
    Q_e(x) = \int Q_{\rm SH}(x^{\prime})\, \mathcal{W}(x, x^{\prime})\, dx^{\prime},
    \label{nonlocaltheory}
\end{equation}
where the kernel $\mathcal{W}(x, x')$ describes the nonlocal contribution of the SH flux from all spatial positions $x'$ to the location $x$, and is defined as

\begin{equation}
    \mathcal{W}(x, x') = \frac{1}{2\lambda(x')} \exp\left[-\frac{X}{\lambda(x')}\right],
    \label{theriticalkernel}
\end{equation}
where the density-weighted distance measure $X$ is given by

\begin{equation}
    X = \frac{1}{n_e(x')} \left| \int_x^{x'} n_e(x'')\, dx'' \right|,
    \label{distance}
\end{equation}
and $\lambda(x')$ represents the effective propagation range of electrons originating from position $x'$:

\begin{equation}
    \lambda(x') \simeq a\, \sqrt{Z + 1}\, \lambda_{\rm free}(x').
    \label{mfp}
\end{equation}
Here, the empirical parameter is taken as $a = 30$ based on prior benchmarks.

From Fig.~\ref{picdata1}, we observe that for a given temperature ratio $R$, simulations with different values of $n_{e0}$ and $L$ - but with the product $n_{e0}L\rm ln\Lambda$ held constant to preserve the same value of $L_T/\lambda_{\rm free}$ for the fixed central temperature - yield the same maximum normalized heat flux $Q_{\rm e,m}/Q_{\rm free}$. 
Note that in our simulation setup, $L_T / \lambda_{\rm free} \sim n_{e0} L \ln\Lambda (R+1)/(R-1)$. Moreover, all such cases exhibit consistent trends. Specifically, when $L_T/\lambda_{\rm free} > 500$, corresponding to the local transport regime, the normalized heat flux approaches the classical SH limit (The SH limit is exceeded within the classical regime at higher $R$ values, attributable to the presence of hotter electrons originating from the hot side). In this regime, the LMV nonlocal theory also agrees well with the simulation results because the kernel $\mathcal{W}(x,x')$ effectively reduces to a Dirac delta function, and therefore recovers local heat flux.

As the system transitions into the nonlocal regime ($L_T/\lambda_{\rm free} < 500$), a clear suppression of the heat flux is observed with respect to the SH model, i.e., $Q_{\rm e} < Q_{\rm SH}$. Notably, the simulation results show good agreement with the LMV model in moderately nonlocal conditions. However, as the degree of nonlocality increases, particularly in cases with lower temperature ratios $R$, the LMV model begins to break down and significantly underestimates the heat flux simulated by the PIC. 

In Fig.~\ref{picdata2} we show that the LMV model fails to accurately reproduce the spatial profile of the heat flux even under moderately nonlocal conditions. In Figs.~\ref{picdata2}(a) through \ref{picdata2}(c), the background electron density is varied from $10^{21}\,\mathrm{cm}^{-3}$ to $10^{25}\,\mathrm{cm}^{-3}$, while the temperature ratio is held constant at $R = 5$, and the characteristic gradient scale length is fixed at $L_T\sim 195$. As a result, the transport parameter varies across the cases: $L_T/\lambda_{\rm free}\approx 83$ in Fig.~\ref{picdata2}(a), $L_T/\lambda_{\rm free}\approx 222$ in Fig.~\ref{picdata2}(b), and $L_T/\lambda_{\rm free}\approx 2500$ in Fig.~\ref{picdata2}(c). In Fig.~\ref{picdata2}(c), the simulation results show excellent agreement with both the SH and LMV models, indicating a return to local transport behavior in the high-collisionality regime. However, in Figs.~\ref{picdata2}(a) and \ref{picdata2}(b), significant deviations are observed in the spatial shape of the heat flux when compared to both models. 

\begin{figure}
    \centering
    \includegraphics[width=0.95\linewidth]{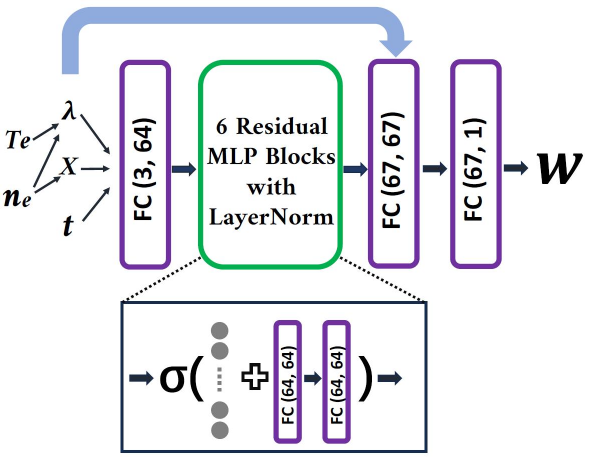}
    \caption{Schematic of \textit{LINN} used to learn the kernel function $\mathcal{W}$. The input features $(\lambda, X)$ are derived from the density and temperature profiles, along with the time coordinate $t$ are first projected into a 64-dimensional latent space via a FC layer followed by LayerNorm and GELU. The core of the network consists of six residual MLP blocks (shown enlarged below), each composed of two FC layers with LayerNorm, GELU, and a skip connection. A high-level skip connection (light-blue arrow) reintroduces the original input vector via concatenation, enabling the model to preserve and leverage key physical features. The resulting 67-dimensional representation is passed through additional FC layers to produce a positive scalar output via Softplus, yielding the kernel weight $\mathcal{W}$.}
    \label{conditionalNN}
\end{figure}

\begin{figure*}
    \centering
    \includegraphics[width=0.95\linewidth]{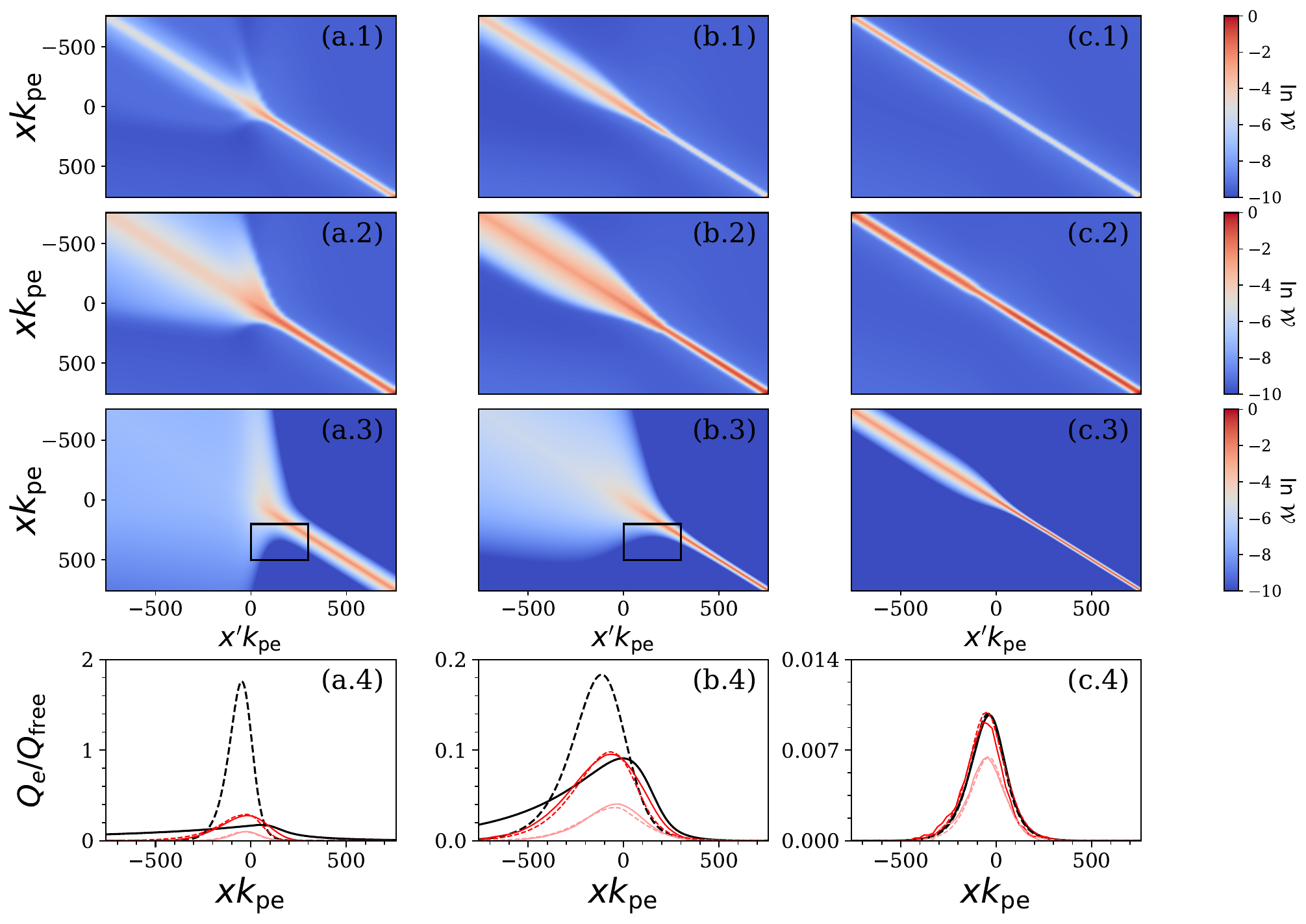}
    \caption{
Comparison of heat flux kernels and resulting heat fluxes for three representative test cases not seen during training. Panels (a)–(c) correspond to cases with increasing collisionality: (a) $n_{e0} = 2.5 \times 10^{20}~\mathrm{cm^{-3}}$, $R = 5$, $L_{\rm T}/\lambda_{\rm free}\approx 33$; (b) $n_{e0} = 7.5 \times 10^{21}~\mathrm{cm^{-3}}$, $R = 5$, $L_{\rm T}/\lambda_{\rm free}\approx 333$; (c) $n_{e0} = 10^{25}~\mathrm{cm^{-3}}$, $R = 2$, $L_{\rm T}/\lambda_{\rm free}\approx 5000$, representing the transition from strongly nonlocal to local transport. For each case, rows (1) and (2) show NN-predicted kernels at the early ($t = t_{\rm sa}/4$) and saturated ($t = t_{\rm sa}$) stages, where $t_{\rm sa}$ denotes the saturation time. Row (3) displays time-independent LMV theoretical kernels. Row (4) presents heat flux comparisons: SH (black dashed), LMV (black solid), and NN predictions (dashed red: $t = t_{\rm sa}$, dashed light red: $t = t_{\rm sa}/4$) against ground truth (solid red: $t = t_{\rm sa}$, solid light red: $t = t_{\rm sa}/4$).
}
    \label{comparisonkernel}
\end{figure*}

The complex and often unpredictable behavior of heat transport in this regime underscores the limitations of traditional theoretical models. This motivates the adoption of data-driven approaches that can capture the underlying physics more accurately. In particular, modern neural networks provide a powerful and flexible framework for learning the transport dynamics directly from high-fidelity simulation data~\cite{miniati,fno,Ingelsten,ucla}, without the need for empirical closures or restrictive analytical approximations.

Motivated by the structure of nonlocal transport theory as expressed in Eq.~(\ref{nonlocaltheory}) and SNB model, where the kernel plays a central role in convolving the local SH heat flux across space, this work proposes a data-driven framework for learning such kernels 
using a neural network (NN) architecture. Drawing inspiration from the physical definitions of $X$ and $\lambda$ in Eqs.~(\ref{distance}) and (\ref{mfp}), the NN is trained to learn the mapping from $(X, \lambda)$ to the kernel function $\mathcal{W}$. To facilitate a dimensionless and numerically stable learning process, the input $\lambda$ is normalized by the spatial grid size $dx$, and the $dx$ within the density integration ($X$) is removed, aligning naturally with the discrete form of Eq.~(\ref{nonlocaltheory}). Furthermore, recognizing that heat transport is inherently dynamic, the temporal coordinate (i.e., the time label, which is normalized by the saturation time corresponding to different samples) is introduced as a third input variable, \textcolor{black}{enabling the model to capture the dynamic transport kernel conditioned on the given $X$ and $\lambda$}. Since the LMV model knowledge is embedded into the neural network, we refer to it as the \textit{LMV-Informed Neural Network (LINN)} hereafter. Then the heat flux can be calculated by
\begin{equation}
    \mathbf{Q}_e(x,t) = \sum_{x'} \mathcal{W}(\raisebox{-2em}{\includegraphics[height=4.5em]{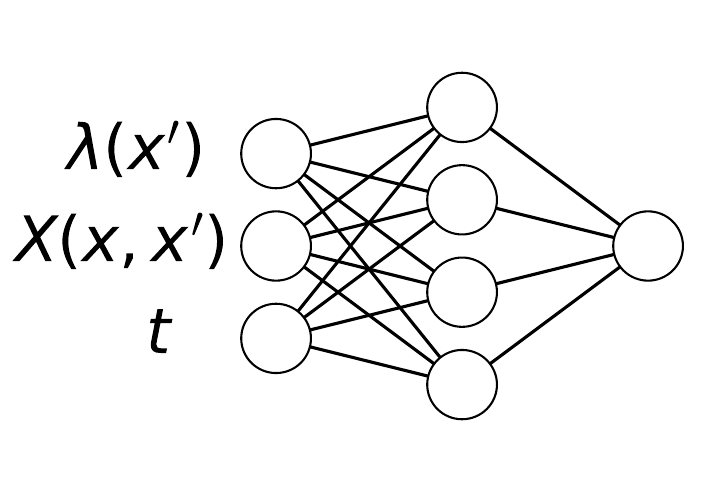}}) \cdot \mathbf{Q}_{\rm SH}(x').
    \label{nonlocalNN_discrete}
\end{equation}

The architecture of our machine learning model, including its inputs and outputs, is illustrated in Fig.~\ref{conditionalNN}. The inputs — $\lambda$ and $X$, computed from the given density and temperature profiles, along with the time coordinate $t$ — are first passed through a fully connected (FC) layer that projects the input into a 64-dimensional latent space, followed by LayerNorm and GELU activation. The core of the network comprises six residual blocks, each consisting of two FC layers with interleaved LayerNorm and GELU activations, wrapped in a residual skip connection (an example block is shown at the bottom of the figure). To preserve and emphasize the original physical input features, a skip connection (light-blue arrow) introduces the raw input vector directly into the post-residual output via concatenation. This combined representation (67-dimensional) is further processed through another FC layer with LayerNorm and GELU, and finally mapped to a non-negative scalar weight using a fully connected output layer with Softplus activation. This architecture balances expressive feature transformation with the retention of physically meaningful input structure, making it well-suited for learning structured kernel functions $\mathcal{W}$ in nonlocal models (see Supplementary Material for details on the NN architecture and model performance).

All $829 \times 2$ samples (including their mirrored counterparts, as the underlying physics exhibit mirror symmetry) are partitioned into training, validation, and test sets using a $70{:}15{:}15$ ratio. Upon constructing the model, we compare the predictions of \textit{LINN} with those of the SH and LMV models, as shown in Fig.~\ref{comparisonkernel}. In Fig.~\ref{comparisonkernel}(a), the central background density is $n_{e0} = 2.5 \times 10^{20}~\mathrm{cm^{-3}}$, with a temperature ratio $R = 5$, and a central value of $L_{\rm T}/\lambda_{\rm free}\approx 33$. Figure~\ref{comparisonkernel}(b) presents results for $n_{e0} = 7.5 \times 10^{21}~\mathrm{cm^{-3}}$, again with $R = 5$ and $L_{\rm T}/\lambda_{\rm free}\approx 333$. In Fig.~\ref{comparisonkernel}(c), the conditions are $n_{e0} = 10^{25}~\mathrm{cm^{-3}}$, $R = 2$, and $L_{\rm T}/\lambda_{\rm free} \approx 5000$. These three scenarios represent a transition from strongly nonlocal to moderately nonlocal and finally to the local transport regime. It is worth noting that none of these cases were included in the training set. The first (1) and second (2) rows correspond to the NN-predicted kernels at the early stage ($t = t_{\rm sa}/4$) and the saturated stage ($t = t_{\rm sa}$), respectively, where $t_{\rm sa}$ denotes the saturation time specific to each case. In comparison, the third (3) row shows the theoretical LMV kernels independent of time. While in the fourth (4) row, the SH and LMV heat flux are presented by the black dashed and solid lines. Because of the correct prediction of the kernels, the calculated heat flux (light-red dashed lines: $t = t_{\rm sa}/4$; red dashed lines: $t = t_{\rm sa}$) via Eq.~(\ref{nonlocalNN_discrete}) obtained good agreement with the ground truth data (light-red solid lines: $t = t_{\rm sa}/4$; red solid lines: $t = t_{\rm sa}$).      

Physically, in the local transport regime, the heat flux kernel is expected to approach a diagonal matrix, reflecting purely local interactions. However, due to temperature asymmetry, the kernel near the hot side should be slightly broader than on the cold side. This is accurately captured by the NN predictions, as shown in Fig.~\ref{comparisonkernel}(c.1) and (c.2). As the system becomes increasingly nonlocal, the kernel broadens more prominently on the hot side, while remaining relatively sharp on the cold side. This trend is clearly observed in Fig.~\ref{comparisonkernel}(a.1), (a.2) and Fig.~\ref{comparisonkernel}(b.1), (b.2), where the NN predictions remain more confined compared to the broader LMV kernels. Notably, the influence of hot particles on the downstream region should decay as the distance from the heat source increases. However, in Fig.~\ref{comparisonkernel}(a.3) and (b.3), the LMV model exhibits a concave feature (highlighted by the black boxes). In contrast, the NN-predicted kernels in Fig.~\ref{comparisonkernel}(a.2) and (b.2) significantly suppress this nonphysical effect, adhering more faithfully to the expected physical behavior. 
The less smooth structures observed in Fig.~\ref{comparisonkernel}(a.1) and (a.2) result from the fact that within the strongly nonlocal regime, the initial temperature profile would experience a non-negligible evolution, and then convolving the local SH flux, which is calculated from the initial temperature profile, across the whole space is not correct. This suggests that the kernel-based approach breaks down in strongly nonlocal regimes~\cite{LMV}. An alternative model that avoids the use of kernels will be explored in future work.

In summary, we present a nonlocal LMV model informed NN that learns nonlocal heat transport kernels directly from structured physical inputs. By incorporating key physical priors—including normalized mean free path, spatial separation, and time—the model captures essential dynamical features while maintaining interpretability. Unlike classical models such as LMV and SNB, which assume static, time-independent kernels, our framework produces adaptive, space- and time-dependent kernels that more faithfully represent the dynamic nature of nonlocal transport. The learned kernels act as flexible, data-driven operators that respond to different plasma conditions while embedding core physical behavior, offering a natural path toward generalization across regimes. Beyond predictive accuracy, the model provides insight into the spatiotemporal structure of heat transport, demonstrating the potential of machine learning embedded with physics models to uncover complex plasma physics~\cite{stefan1,stefan2,mfluo1,mfluo2} in ways that extend beyond traditional analytical frameworks. Future embedding of these kernels into large-scale hydrodynamic modeling has the potential to enhance predictive capabilities of complex plasma environments, including astrophysical systems and inertial confinement fusion experiments. 
 
The authors thank S. Nathaniel, S. Hüller, T. Fülöp, and C. Riconda for their useful discussions. They are also grateful for the computing resources provided by the STFC Scientific Computing Department’s SCARF cluster. This work was supported by EPSRC and First Light Fusion under the AMPLIFI prosperity partnership, Grant No. EP/X025373/1.

%

\end{document}